\documentclass[fleqn,twoside,twocolumn,nofootinbib]{revtex4} % Specifies the document class %,unsortedaddress
\usepackage{ujp} % \usepackage[cyr]{ujp} for cyrillic, \usepackage[web]{ujp} for web
\begin{document}
\title[Vibrational spectra of berberine]%колонтитул
{Vibrational spectra of berberine and their interpretation~~ by~~ means~~ of~~ DFT quantum-mechanical calculations}%
\author{N. Bashmakova}%1 автор
\affiliation{Taras Shevchenko National University of Kyiv}%институт
\address{64, Volodymyrs'ka Str., Kyiv 01601, Ukraine}%адрес
\email{lns@univ.kiev.ua}%e-mail
\author{S. Kutovyy}%
\affiliation{Taras Shevchenko National University of Kyiv}%
\address{64, Volodymyrs'ka Str., Kyiv 01601, Ukraine}%
\email{lns@univ.kiev.ua}%e-mail
\author{R. Zhurakivsky}
\affiliation{Institute of Molecular Biology and Genetics}%
\address{150, Zabolotnyi Str., Kyiv 03143, Ukraine}%
%\email{}%
\author{D. Hovorun}%
\affiliation{Institute of High Technologies, Taras Shevchenko
National
University of Kyiv}%
\address{2, bd. 5, Academician Glushkov Ave., Kyiv 03022, Ukraine}%
%\email{}%
\author{V.~Yashchuk}%
\affiliation{Taras Shevchenko National University of Kyiv}%
\address{64, Volodymyrs'ka Str., Kyiv 01601, Ukraine}%
\email{lns@univ.kiev.ua}%
\udk{535.37; 535.58} \pacs{33.20.-t, 71.15.Mb,\\[-3pt] 87.15.-v} \razd{}

\setcounter{page}{130}%
\maketitle

\begin{abstract}
Experimental vibrational spectra (Raman and infrared absorption) of
berberine are obtained at room temperature. The vibrational
spectra of berberine are calculated by the DFT method at the
B3LYP/6-311++G(d,p) level. Based on the correlation between experimental
and calculated data, the vibrational spectrum is interpreted
in the frequency range of 800--1700 cm$^{-1}$ in detail. The experimental
and calculated spectra of intramolecular vibrations are found to
correlate closely.
\end{abstract}

\section{Introduction}

Natural alkaloid berberine has been used in medicine for a long time for
treating a number of diseases. Preparations of berberine have been
found to have cytotoxic, bactericidal, and antiviral effects [1].
Moreover, berberine influences the cancer cell metabolism by destroying
cells irreversibly.
%The toxicity of berberine preparations is relatively low, and the cured organism develops immunity against relapse [1,2].
Such activity of the alkaloid is attributed to its ability to
intercalate into the DNA macromolecule by blocking the processes of
replication and transcription. Many articles are devoted to
the interaction between berberine and nuclei acids, see, e.g., reviews
[1,2]. Another important target for antitumour berberine agents is
DNA-topoisomerase: berberine can inhibit topoisomerase, by breaking the
connection between this enzyme and DNA [3].

To clarify which mechanism causes berberine to produce its
therapeutic effect, it is necessary to know the mechanisms by which
it binds to DNA. Earlier, using the Raman spectroscopy method, we
revealed an interesting fact of a resonance interaction of berberine
and DNA vibrations [4, 5]. It was found that, in the DNA-berberine
solution, the intensity of vibrations of both berberine and DNA increases
greatly (by orders as compared to those of the spectra of components) in the
range where their intense spectra overlap (1000--1700
cm$^{-1}$). To explain this interaction, the detailed data about
berberine and DNA vibrational spectra are required.

%Це виглядає як прояв резонансної взаємодії молекул ДНК і берберину, причому саме в діапазоні перекриття інтенсивних спектрів цих молекул (1100-1700 cm$^{-1}$).

The vibrational spectra of berberine were previously investigated
with the use of the Raman [6--9] and IR-absorption spectroscopy methods [7,
9], but no exhaustive interpretation of the obtained spectra was given in
those works. In works [7--9], the Raman spectra of berberine in the range
of 600--1800 cm$^{-1}$ obtained by the SERS and SSRS
(surface-enhanced and shifted-subtracted Raman spectroscopy) methods
were presented.

This work continues and develop our study described in [10].

\section{Materials and Methods}

Natural berberine is mainly contained in plants as hydrochloride
or dihydrosulfate, a structural formula of a neutral berberine
molecule is C$_{20}$H$_{19}$NO$_{5}$ (or
C$_{20}$H$_{18}$NO$_{4}$(OH)). A structural formula of a berberine
cation (as usually investigated) is [C$_{20}$H$_{18}$NO$_{4}]^{+}$
(see Fig.~1). The quantum-mechanical investigations of
the berberine structure were carried out in [11, 12]. The results of
studies of the spatial structure of berberine cation isomers
[12] demonstrate that the berberine cation that consists of
hexamerous planar rings has an almost planar frame structure.
A deviation from this planar structure is observed only in a partially
saturated C ring. It conforms to the shape of a half-chair, in which
C7 and C8 atoms of the C ring deviate appreciably from the plane of B and D
rings.

The microcrystalline powder of berberine has been studied
(berberine hydrochloride, ``Alps Pharmaceutical'', Japan, 407
g/Mole). Since berberine absorbs in the range to 550 nm [13], a
He-Ne laser emission ($\lambda=6328$ \AA) has been used for
the excitation of Raman spectra; the real power on a specimen was
$\sim$\,10~mW. Some spectra were obtained by using a Kr-laser
($\lambda=6471$~\AA) with a power $\sim$\,60--80 mW on a specimen.
Raman spectra have been measured at room temperature with the
$45^{\circ}$-reflection geometry using an optical setup made on the
basis of a double-grating monochromator DFS-24 with a resolution of
1.5~cm$^{-1}$. A laser beam was focused on a specimen with a
cylindrical lens to obtain an elongated irradiated area parallel to
the entrance slit of the spectrometer. To improve the
signal-to-noise ratio, the time accumulation at the point was
$\sim1.6$~s (multipass mode). Spectra were recorded in the range of
40--4000 cm$^{-1}$.

The IR-absorption spectra of berberine (microcrystalline powder)
were recorded with a Nicolet NEXUS-470 Fourier spectrometer. We used
the ATR (\emph{Attenuated Total Reflectance}) technique.  To improve
the signal-to-noise ratio, a signal accumulation regime was used
(128 scans). Spectra were recorded in the range of 400--4000
cm$^{-1}$ with a resolution $\leq 1$ cm$^{-1}$. In the resulting spectra,
Fresnel losses on the input and output surfaces of specimens were
allowed for. In addition, to obtain the absorption spectra, the dependence
of the penetration depth of evanescent waves into the sample on
their wavelength is considered according to the relation [14]
%1
\begin{equation}
d_{p}=\frac{\lambda}{2\pi n \sqrt{\sin^{2}\theta-(n/n_{s})^{2}}},
\end{equation}
where $n$ and $n_{s}$ are refractive indices of a prism (diamond) and a
specimen, respectively, $\lambda$ is a light wavelength, and $\theta$
is an incident beam angle.

In some cases, experimental spectra were treated, by using the
\emph{Origin} and \emph{PeakFit} programs to specify the data on the
band position, the number of band components, and so on.

The optimized geometry of the berberine cation was calculated by the DFT
method at the B3LYP/6-311++G(d,p) level without any structural
restrictions. At this level, the vibrational (Raman and
IR-absorption) spectra were calculated in the harmonic approximation. It
should be noted that, for the first time in an optimized structure,
the intermolecular hydrogen bond  CH…O  between metoxylic groups
OCH$_{3}$ of berberine was found [10]. The existence of the bond was
determined based on the presence of a critical point of (3,--1)
type on the electron density distribution by the АІМ method [15].
The binding energy was found to be 3.12 kcal/mole. The DFT
calculations were performed using the ``\emph{Gaussian03}'' program
package for \emph{Win32} [16] granted by the ``\emph{Gaussian}''
corporation.

%Fig. 1
\begin{figure}% figure* for wide figure, [h] [!] to change the placement
\includegraphics[width=\column]{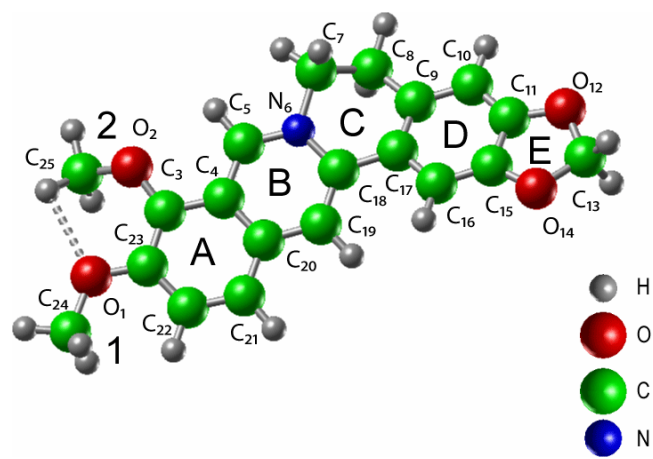}
\caption{Structural formula of a berberine cation (calculation
by the DFT method at the B3LYP/6-311++G(d,p) level)  }
\end{figure}

Because the DFT method usually overestimates frequencies, a scaling
factor of 0.985 was used for the comparison of calculated and
experimental data. This correction was necessary due to errors in
the calculation of interatomic interactions as a result of the limited basic
functions set [17]. The value 0.985 provides the best correspondence
between calculated and experimental data. The scaling factor was
determined by Raman spectra -- as it turned out, the shapes
of calculated and experimental spectra was more
similar for Raman spectra, and the correspondence between
calculated and experimental frequencies was practically unambiguous. In our case, the
scaling factor is close to 1 indicating that the used basic set is
sufficient for the studied molecule.\looseness=1

In addition, it is known that the \emph{Gaussian} set of programs deals with
Raman activities $S_{i}$, not intensities $I_{i}$ [18, 19].
Therefore for comparing with the experiment, the Raman activities
$S_{i}$ were corrected by the following relationship to obtain Raman
intensities [18]:
%2
\begin{equation}
I_{i}=\frac{C(\nu_{0}-\nu_{i})^{4}S_{i}}{\nu_{i}[1-\exp(-hc\nu_{i}/kT)]}.
\end{equation}
Here, $C$ is a constant, $\nu_{0}$ is the laser
excitation line frequency, $\nu_{i}$ is the vibrational frequency, $h, k, c,
T$ are Planck and Boltzmann constants, speed of light, and temperature
in degrees Kelvin, respectively.

The values of frequencies and relative intensities of spectral lines with regard for the
aforementioned corrections and the detailed interpretation of vibrations are
presented in Table.

%Fig. 2
\begin{figure}% figure* for wide figure, [h] [!] to change the placement
\includegraphics[width=\column]{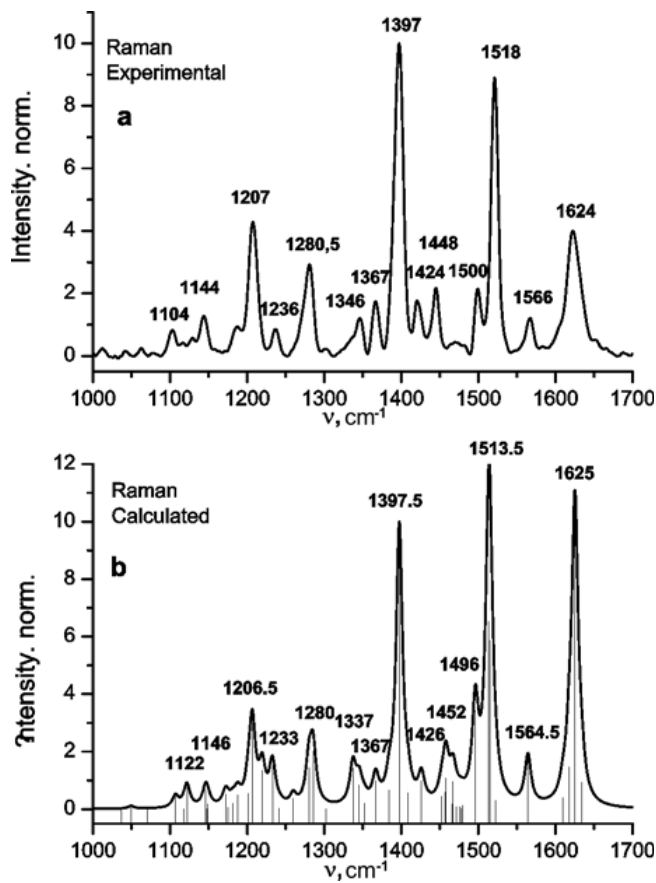}
\caption{Raman spectra of berberine in the range of 1000--1700
cm$^{-1}$: experimental, $\lambda _{\rm ex}=6328$~{\AA} ({\it a}),
and calculated ({\it b})  }
\end{figure}

\section{Results and Discussion}

The berberine crystal
(C$_{20}$H$_{18}$NO$_{4}^{+}$Cl$^{-}$4H$_{2}$O) symmetry is
triclinic (\emph{P1}, $z=2$) [20]. Due to a low symmetry, there are no degenerated vibrations in
vibrational spectra of berberine.
The correspondence between the ``crystalline'' and
``molecular'' modes is unambiguous. The most essential difference
between the spectra of free and crystalline berberine had to reveal
itself in the range of very low frequencies (the so-called external
vibrations, there are three of them at $z=2$. In the range of middle
and high frequencies (internal vibrations), the influence of a
crystal structure is very weak. Both duplication in the number of
vibrations (because of $z=2$) and some shifting of the vibrational
bands take place.

%Fig. 3
\begin{figure}% figure* for wide figure, [h] [!] to change the placement
\includegraphics[width=\column]{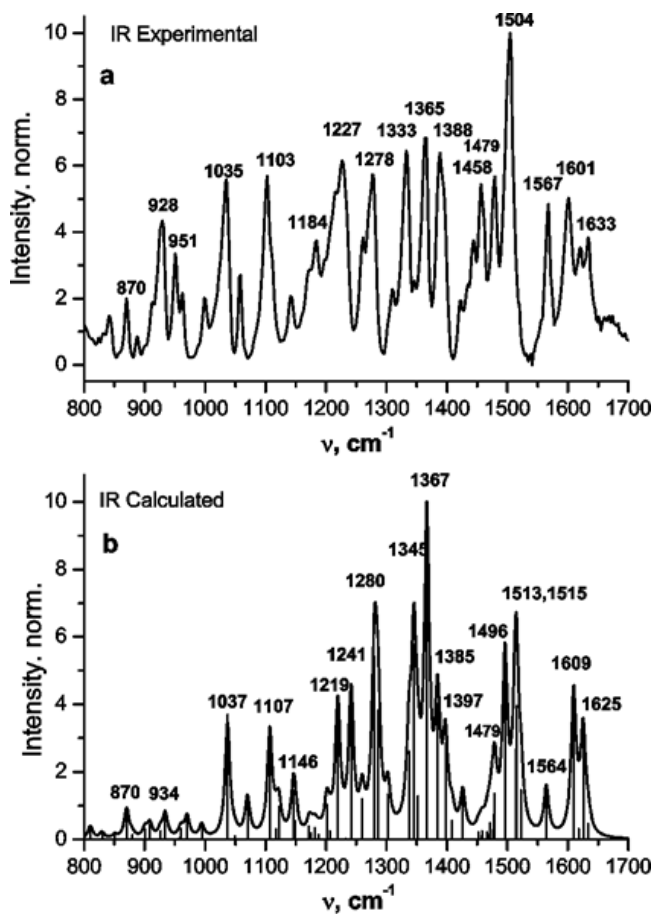}
\caption{IR-absorption spectrum of berberine in the range of
800--1700 cm$^{-1}$: experimental ({\it a}), and calculated ({\it
b}) }
\end{figure}

Therefore, we believe that the comparison of experimental vibrational
spectra of microcrystalline berberine with calculated spectra of the
berberine cation in the actual range of 600--1800 cm$^{-1}$ was
sufficiently correct, which was confirmed by the following analysis.
Experimental and calculated Raman and IR-absorption spectra in the
range of 800--1700 cm$^{-1}$ are presented in Figs.~2 and 3,
respectively. The frequency values -- both experimental and
calculated using the scaling factor -- are presented in Table. The
relative intensities of Raman lines were corrected by (1). In this
case, we are interested more in the range of 1000--1700 cm$^{-1}$.
This is because berberine vibrations are the most
intense in this range, and there a resonance interaction of the
berberine and DNA vibrations (increasing by orders) is observed in
the Raman spectrum of a berberine-DNA water solution [4,~5].

%Табл. 1
\begin{table*}[!]
\noindent\caption{Experimental and calculated frequency values, as
well as the interpretation of vibrations. The frequency scaling factor
is 0.985, calculated Raman intensities were corrected by
relationship (2), experimental IR-intensities --  by
(1)}\vskip3mm\tabcolsep2.6pt

\noindent{\footnotesize\begin{tabular}{c c c c c c c c l}
 \hline \multicolumn{5}{c}
{\rule{0pt}{9pt}Experimental Data} &
\multicolumn{3}{|c}{Calculations  }&
\multicolumn{1}{|c}{Interpretation of vibration modes}\\%
\cline{1-5} \multicolumn{1}{c}{[6] Ram}&\multicolumn{1}{|c}{[9] Ram
}&\multicolumn{1}{|c}{[7] IR}& \multicolumn{2}{|c}{Our data}&
\multicolumn{3}{|c}{with corrections}&
\multicolumn{1}{|c}{}\\\cline{4-8}%
\multicolumn{1}{c}{}&\multicolumn{1}{|c}{}&\multicolumn{1}{|c}{}&
\multicolumn{1}{|c}{Ram}&\multicolumn{1}{|c}{IR}&
\multicolumn{1}{|c}{$\omega$, cm$^{-1}$}&\multicolumn{1}{|c}{$I_{\rm
Ram}$}&\multicolumn{1}{|c}{$I_{\rm IR}$}&
\multicolumn{1}{|c}{}\\%
\hline%
 --    & 731  &    -- &   727 &       --  &   728  & (2.0)  &  (0.1) &  p(СН$_{3}$); p,$\tau$(СН$_{2}$) r.С; f(СН$_{2}$) r.E;  \\
 &&&&&&&&in-plane $\nu-\delta$ of all rings; $\nu,\delta$(CNC);
 $\delta$(СOС);\\
$\ldots$&       &       &$\ldots$&     &$\ldots$&       &  &
$\ldots$ \\%
 --   & --     & --     &--  & vw813 &810    & (vw) &
(0.4)  &  out-of-plane of atoms H,C r.A,B (H--C22,21,19); others weak \\%
 --   & 834   & --     &--  & vw832 &830    & (0.4) & (0.2) &
$\nu$(O--CH$_3$); very strong $\nu$-$\delta$ r.D,E; weaker
$\nu$-$\delta$ r.A,B,C;  \\%
--   & 842   & --     &843    & 842   &851    & (vw)  & (0.1) & out-of-plane of atoms H,C r.D (H--C10,16) --  \\%
--   & --     & --     &--  & 870   &870    & (vw)  & (1.0) &
antiphase (851) and cophase (870); p(C8H2) r.C;  \\%
&&&&&&&&  weak $\nu$-$\delta$ of all rings; \\%
 --   & --     & --     &--  & 888
&879    & (0.2) & (0.1) & $\nu$(O--CH$_3$);   p(CH$_2$) r.C; strong
$\nu$-$\delta$ of all rings;   \\%
 --   & --     & 898 &-- & 900 &901
& (vw)  & (0.3) & out-of-plane of atoms H,C19 r.B (H--C19);
p(CH$_2$)
r.C; \\&&&&&&&&  others weak;   \\%
 --   & --     & -- &--  & 912 &908    &
(0.1)     & (0.5) & $\nu$(O--CH$_3$); p(CH$_2$)
r.C;   strong O--C--O r.E;     \\%
&&&&&&&&in-plane $\nu$-$\delta$ of all rings; \\%
 --   & --     & --     &--  & 928   &926    & (vw)  & (0.3) &
out-of-plane of atoms H,C5 r.B (H--C5);   others very weak;   \\%
--   & --     & --     &--  & -- &934    & (0.2)     & (0.4) &
p(CH$_2$) r.C;   $\delta$ r.C,D;   $\nu$-$\delta$ r.E (O--C--O very strong);\\%
--   & --     & --     &--  & -- &957    & (vw)  & (vw)  &
out-of-plane of atoms H r.A,B;   others very weak;   \\%
--   &-- & --     &--  & 951   &960    & (vw)  & (0.3) &
$\delta$(CH$_3$(1)); p(CH$_3$(2));   p,f(CH$_2$) r.C;   f(CH$_2$)
r.E;  \\%
&&&&&&&&   strong in-plane  $\nu$-$\delta$ of all rings; \\%
 -- & -- & --
&970 & 963   &970    & (0.3) & (0.7) & weak out-of-plane r.А
(С21--H,
C22--H);\\%
 --   & --     & 1002  &--  & 1000  &994    & (vw) & (0.4)
& p(CH$_3$(2));  $\nu$(O1--CH$_3$);  f,$\tau$(CH$_2$) r.C;    \\%
&&&&&&&&f(CH$_2$) r.E;  strong $\nu$-$\delta$ all rings;\\%
 -- & --     & 1035 &--
& 1035  &1037   & (vw)  & (3.7) & strong
$\nu$-$\delta$(O--CH$_2$--O) r.E and $\nu$-$\delta$ r.D;  \\%
&&&&&&&&   out-of-plane pf(CH$_2$) r.C;   weak of other rings; \\%
 --   & 1044  & -- &--  &
-- &1050   & (vw)  & (0.1) & u(CH$_3$);   out-of-plane
pf(CH$_2$) r.C;   f(CH$_2$) r.E;   $\nu$-$\delta$ of all rings;   \\%
-- & 1067  & 1065  &--  & 1058  &1070   & (vw)  & (1.3) & u(CH$_3$);
$\nu$(O--CH$_3$);   out-of-plane p(CH$_2$) r.C;     \\%
&&&&&&&&f(CH$_2$) r.E; $\nu$-$\delta$ of all rings; \\%
 --   & --
& -- &1104 & 1103 &1107   & (0.4) & (3.4) & u,p(CH$_3$); f(CH$_2$);
$\nu$(O--CH$_3$), $\nu$(C--OCH$_2$) of both pairs  \\%
&&&&&&&&  of bonds;  in-plane $\nu$-$\delta$ of all rings;   H -- in-plane; \\%
--   & --     & --     &--  & 1114sh&1117    & (vw)  & (0.3) & out-of-plane strong p(CH$_2$) and weak of atoms O r.E;\\%
&&&&&&&&weak $\nu$-$\delta$ of all rings;\\%
 1118    & -- & 1110
&1119   & 1121sh&1122    & (0.8) & (1.0) & u,p(CH$_3$); out-of-plane
pf(CH$_2$) r.E; $\nu$(O--CH$_3$),
$\nu$(C--OCH$_2$),   \\%
&&&&&&&& $\nu$(N--CH$_2$), $\nu$(C7--CH$_2$) r.C; $\delta$ all
rings;\\&&&&&&&&  H -- in-plane and out-of-plane; \\%
 1144    & -- & 1143 &1144 & 1142  &1146   & (0.7) & (1.5) & strong
$\nu$(N-CH$_2$), $\delta$(C5NC18);      \\%
&&&&&&&&$\nu$-$\delta$ r.A,B,C and atoms H in bonds C--H;   r.E
rigid;\\[2mm]%
--   & --     & --     &--  & -- &1148   & (vw)  & (vw)  & strong p(CH$_3$(1)), others weak;    \\%
--   & --     & --     &--  & vw1152&1149   & (0.2) & (0.6) & strong
p(CH$_3$(2)), others weak;    \\[2mm]%
sh   & --     & --     &1174   & 1171  &1172   & (0.5) & (0.4) &  p(CH$_3$); with differ. ampl.;   $\tau$,f(CH$_2$) r.C; $\tau$(CH$_2$) r.E; \\%
--   & --     & --     &--  & -- &1175   & (vw)& (0.2) &       $\delta$-vibrations r.D, mating with $\nu$-vibrations r.C;\\%
--   & --    & --     &--  & -- &1181   & (0.2) & (0.3) &    $\delta$(C7NC18);   $\delta$-vibrations r.A,B; \\%
--   & --     & 1182  &1187.5 cm & 1184 &1187.5 & (0.5) & (0.2) &
(1181 --  strong only p(CH$_3$), 1187- $\tau$(CH$_2$) r.E);\\[2mm]%
 -- & -- &
-- &--  & 1199  &1201   & (0.5) & (1.0) & p(CH$_3$); $\tau$(CH$_2$)
(r.C -- strong, r.E -- weak);   strong $\nu$(N-C7,  \\%
&&&&&&&& N-C18) and $\delta$(C17C9C8 );   weak $\nu$ of all rings; \\%
1203    & 1206  & --     &1206&vw1207sh  &1206.5  & (3.0) & (0.2) & p(CH$_3$);   $\tau$(CH$_2$) (r.C -- strong, r.E -- weak);  \\%
    &   &   &   &   &       &   &       &    $\nu$-$\delta$ of all rings; atoms H of rings -- in-plane;  \\%

 \hline
\end{tabular}}
\end{table*}

%Табл. 2
\begin{table*}[!]
\noindent\caption{Experimental and calculated frequency values, as
well as the interpretation of vibrations. The frequency scaling factor
is 0.985, calculated Raman intensities were corrected by
relationship (2), experimental IR-intensities --  by (1)
(Continuation)}\vskip3mm\tabcolsep3.3pt

\noindent{\footnotesize\begin{tabular}{c c c c c c c c l}
 \hline \multicolumn{5}{c}
{\rule{0pt}{9pt}Experimental Data} &
\multicolumn{3}{|c}{Calculations  }&
\multicolumn{1}{|c}{Interpretation of vibration modes}\\%
\cline{1-5} \multicolumn{1}{c}{[6] Ram}&\multicolumn{1}{|c}{[9] Ram
}&\multicolumn{1}{|c}{[7] IR}& \multicolumn{2}{|c}{Our data}&
\multicolumn{3}{|c}{with corrections}&
\multicolumn{1}{|c}{}\\\cline{4-8}%
\multicolumn{1}{c}{}&\multicolumn{1}{|c}{}&\multicolumn{1}{|c}{}&
\multicolumn{1}{|c}{Ram}&\multicolumn{1}{|c}{IR}&
\multicolumn{1}{|c}{$\omega$, cm$^{-1}$}&\multicolumn{1}{|c}{$I_{\rm
Ram}$}&\multicolumn{1}{|c}{$I_{\rm IR}$}&
\multicolumn{1}{|c}{}\\%
\hline%
--   & --     & 1230  &1222   & 1227  &1219.5 & (1.3) & (4.2) & u(CH$_3$) (1-weak);    \\%
&&&&&&&&$\tau$(CH$_2$) r.C;   f(CH$_2$) r.E;  $\nu$(O(1.2)-C);
$\delta$(OCO) r.E;\\&&&&&&&&   $\nu$-$\delta$ of all rings\\%

1235    & 1237  & --     &1236   & 1234sh&1233    & (1.5) & (0.1) &
(1219 -- decreasing from A
to D;  1233 -- conversely);   \\[2mm]%
--   & --     & -- &--& vw1246sh &1241.5 & (vw)  & (4.5) &
p(CH$_3$); $\tau$(CH$_2$) r.C; weak f(CH$_2$) r.E;   $\nu$-$\delta$
all rings and O,N;
\\%
 sh   & --     & --     &1257sh & 1261   &1260   & (0.4) &
(1.2) & weak p,u(CH$_3$);   $\tau$(CH$_2$) r.C;   f(CH$_2$) r.E;
  \\%
&&&&&&&&$\nu$-$\delta$ of all rings (r.B,C,D,E -- strong); \\%
 1276 &1280  & 1271  &1280.5 & 1278  &1280   & (1.4) & (5.4) & p,u(CH$_3$);
weak $\tau$,f(CH$_2$) r.C;   weak f(CH$_2$) r.E; \\%
&&&&&&&&$\nu$-$\delta$ of all rings (r.A,B -- strong); \\%
 --   & -- &1301  &1295   & -- &1285.5 & (2.0) & (3.8) & p,u(CH$_3$); strong
f(CH$_2$) r.C; weak $\delta$(CH$_2$) r.E;
\\%
&&&&&&&&$\nu$-$\delta$ of all rings (r.C,D,E -- strong);\\%
 --   & --
& --     &1303 & 1309  &1302   & (vw)  & (1,3) & u(CH$_3$);
$\tau$(CH$_2$) r.C, weak f(CH$_2$) r.E; \\%
&&&&&&&& $\nu$-$\delta$ of all rings (r.A,B-strong);\\[2mm]%
sh   & --     & 1331  &1333   & 1333  &1337   & (1.5) & (2.6) &  u(CH$_3$);   f,f or f,$\tau$(CH$_2$) r.C, f(CH$_2$) r.E;  \\%
1342    & 1340  & --     &1346   & 1346  &1345.5 & (0.8) & (5.8) &  $\nu$-$\delta$ of all rings (1337 -- strong, 1352 -- weak); \\%
--   & --     & --     &--  & -- &1352   & (0.2) & (1.3) & \\[2mm]%
1361    & 1367  & 1364  &1367   & 1365  &1367   & (1.0) & (10)  & p(CH$_3$);   f,$\tau$(CH$_2$) r.C, f(CH$_2$) r.E;  \\%
--   & --     & 1390  &1385sh & 1388   &1384.5 & (0.7) & (4.0) &
$\nu$-$\delta$ vibrations of all rings (1384 -- commensurable,    \\%
&&&&&&&&1367 -- decreasing from A to E);\\[2mm]%
 1397    & 1397  & --     &1397   &
1397sh&1397.5  & (10)  & (2.8) & p(CH$_3$), p(CH$_2$);   in-plane
$\nu$-$\delta$-vibrations of atoms\\&&&&&&&& C,N,O,H of  all rings;   \\%
-- & -- & --     &1410   & --     &1408.5 & (0.6) & (0.6) & strong
f(CH$_2$) r.E, weaker -- r.C; weak u(CH$_3$);\\&&&&&&&&
$\nu$-$\delta$
vibrations of all rings;   \\%
1424    & 1425  & 1424  &1424 & 1422  &1426   & (1.0) & (1.3) &
u(CH$_3$)-strong (2);   f(CH$_2$); atoms O -- almost
immobile;\\%
&&&&&&&& strong $\nu$-$\delta$ vibrations of all rings;\\[2mm]%
1449    & 1447  & --     &1448   & 1444  &1452   & (0.5) & (0.2) & u(CH$_3$): 1457-strong (1);   $\delta$(CH$_2$); \\%
&&&&&1457&(0,6)&(0,1)&$\nu$-$\delta$ vibrations of all rings\\
%start of row
--   & --     & --     &1460   & 1457  & 1458     & (1.1) & (0.2)&(1452-strong, 1457-weak), r.C,D,E (1458); \\%
sh   & --     & --     &1466   & 1466sh&1466    & (0.2) & (0.2) &
$\delta$(CH$_3$): 2-strong, 1-weak;   \\%
--   & --     & --     &1472   & -- &1467.5 & (1.0) & (0.1) & u(CH$_3$);   $\delta$(CH$_2$) r.C;  $\nu$-$\delta$ vibrations r.A,B,C;   \\
 --  & --     & --     &--  & -- &1471   & (vw) & (0.5) & \\[2mm]%
sh   & --     & --     &1475b& --  &1476   & (0.1) & (0.3) & with differ. ampl. $\delta$(CH$_3$);      \\%
--   & --     & --     &--  & -- &1478   & (vw)  & (0.7) & weak vibrations adjacent r.A; \\%
 --  & 1481  & --     &1487   & 1479  &1479   & (0.1) & (1.4) & \\[2mm]%
1501    & 1499  & 1506  &1500   & 1504  &1496   & (3.4) & (5.6) &
u(CH$_3$); $\delta$(CH$_2$) r.E and C;  $\nu$-$\delta$ vibrations
r.D
(strong),\\&&&&&&&& r.A,B (weak);   \\[2mm] %
1518    & 1520  & --     &1518   & 1511sh&     1513.5  & (6.6) & (2.7)&u,$\delta$(CH$_3$);   $\delta$(CH$_2$); in-plane $\nu - \delta$-vibrations with\\
    &&&&&1515    & (5.9) & (4.0)&differ. ampl. of atoms C,N,O,H of all rings;\\[2mm]%
 --   & --     & --     &1532sh & 1520sh&1522.5 & (0.3) & (1.5) &  strong $\delta$(CH$_2$) r.E, weak $\nu$-vibration r.E,D,\\&&&&&&&& weak of others rings;   \\%
1568    & 1569  & 1558  &1566   & 1567  &1564.5 & (1.8) & (1.6) &
u(CH$_3$): strong (2), weak (1);   strong in-plane $\nu$-$\delta$
vibrations    \\%
&&&&&&&& r.A,B  (including C18-N and gr.C7H2 r.C);\\%
 \hline
\end{tabular}}
\end{table*}

%Табл. 3
\begin{table*}[!]
\noindent\caption{Experimental and calculated frequency values as
well as interpretation of vibrations. The frequency scaling factor
is 0.985, calculated Raman intensities were corrected by
relationship (2), experimental IR-intensities --  by (1)
(Continuation)}\vskip3mm\tabcolsep5.2pt

\noindent{\footnotesize\begin{tabular}{c c c c c c c c l}
 \hline \multicolumn{5}{c}
{\rule{0pt}{9pt}Experimental Data} &
\multicolumn{3}{|c}{Calculations  }&
\multicolumn{1}{|c}{Interpretation of vibration modes}\\%
\cline{1-5} \multicolumn{1}{c}{[6] Ram}&\multicolumn{1}{|c}{[9] Ram
}&\multicolumn{1}{|c}{[7] IR}& \multicolumn{2}{|c}{Our data}&
\multicolumn{3}{|c}{with corrections}&
\multicolumn{1}{|c}{}\\\cline{4-8}%
\multicolumn{1}{c}{}&\multicolumn{1}{|c}{}&\multicolumn{1}{|c}{}&
\multicolumn{1}{|c}{Ram}&\multicolumn{1}{|c}{IR}&
\multicolumn{1}{|c}{$\omega$, cm$^{-1}$}&\multicolumn{1}{|c}{$I_{\rm
Ram}$}&\multicolumn{1}{|c}{$I_{\rm IR}$}&
\multicolumn{1}{|c}{}\\%
\hline%
-- &  --  & 1600  & 1613  & 1601  &1609.5 & (0.4) & (4.5) & in-plane $\nu$-$\delta$ vibrations of atoms C,O,H,N of all rings; \\%
sh&   -- &  --    &1623   &  --    & 1618  & (1.5) & (0.3) &    weak u,p(CH$_3$);   $\tau$(CH$_2$) {r.C}, f(CH$_2$) r.E; \\%
1626    & 1622  & 1629  &  1624 &  1620 & 1625  & (10.7)& (3.3) & 1609: strong of all rings;   O -- almost immobile;\\%
  & & & & & & & &   1618: strong vibrations r. А,D; N -- immobile; \\
  & & & & & & & & 1625: strong vibrations -- r.D, N -- vibrates;   \\[2mm]
 sh     & 1633  &  --    &  1635 &  1633 & 1634  & (0.9) & (0.5) &  in-plane $\nu-\delta$ of atoms С,N,O,Н of all rings  \\%
 &&&&&&&&(strong -- r.В,С,D, weak -- gr. СН$_{2}$; gr. ОСН$_{3}$ --
 immobile);\\%
\hline
\end{tabular}
\vskip1mm \noindent{N o t e s}: In brackets near frequency values
are the relative intensity of lines.  (The intensities of the modes
at 1397 cm$^{-1}$ in Raman\\ spectra and 1367 cm$^{-1}$ in
IR-absorption spectra count as 10. Modes with relative intensity
less than
0.1 are designated as ``vw'', for\\ ``very weak'').~~~~~~~~~~~~~~~~~~~~~~~~~~~~~~~~~~~~~~~~~~~~~~~~~~~~~~~~~~~~~~~~~~~~~~~~~~~~~~~~~~~~~~~~~~~~~~~~~~~~~~~~~~~~~~~~~~~~~~~~~~~~~~~~~~~~~~~~~~~~~~~~~~~~~~~~~~~~~~\\
{A b b r e v i a t i o n s}:~ \emph{b}:~ band; \emph{sh}:~ band
shoulder; {r.A,B,C,D,E}: A, B, C, D and E rings, respectively;
\emph{weak}'' and ``\emph{strong}'' denote vibrations with small and
large displacements, respectively. Vibrations: $\nu$ -- stretching,
$\delta$ -- deformation (for the CH$_{2}$-groups, the same as~
bending);\, $\tau$:\, --\, torsion;\, $\nu-\delta$:\, a\, part\, of~
C\, atoms\, in\, a\, ring\, performs\, $\nu$-vibrations,\, and\,
the\, others~ perform~ $\delta$-vibrations;~ for CH$_{2}$-groups --
\emph{f}: wagging; \emph{p}: rocking (atoms vibrate in one plane);
for CH$_{3}$-groups -- \emph{u}: umbrella (symmetric
$\delta$-vibrations); \emph{p}: rocking (atoms vibrate in parallel
planes); for $\tau$, \emph{f}, \emph{p} -- vibrations preserve the
rigidity of CH$_{3}$ and CH$_{2}$ groups.\hfill}\vspace*{-2mm}
\end{table*}

Unfortunately, the high frequency region of 3000--3300 cm$^{-1}$ was
not applicable to the comparative analysis: Raman spectra have very
small intensities in this region, and IR-absorption spectra are only
loosely correlated with calculations.

The calculated spectrum consists of $123 (3\times43-6)$
non-degenerate vibrations of a berberine cation
[C$_{20}$H$_{18}$NO$_{4}]^{+}$, 105 of which lie in the range of
20--1700 cm$^{-1}$ (61 of which are in the actual range of 800--1700
cm$^{-1}$), other 18 ones fall in the range of 3000--3300 cm$^{-1}$.
Low-range frequencies up to 720 cm$^{-1}$ correspond to out-of-plane
vibrations of rings and associated groups. In-plane vibrations of
rings appear beginning with 720 cm$^{-1}$; in particular, the only
sufficiently intense 727-cm$^{-1}$ Raman mode  outside of the range
of 1200--1700 cm$^{-1}$. In-plane vibrations are characteristic of
the entire range from 720 to 1700 cm$^{-1}$. The high-frequency
region of 3000--3300 cm$^{-1}$ consists of the modes corresponding
to vibrations of C--H-bonds with weak displacements of other atoms,
mostly C.

In Fig. 2,{\it a}, the experimental Raman spectrum of berberine
(microcrystalline powder) in the range of 1000--1700 cm$^{-1}$ at
the excitation by the 6328-{\AA} line is presented. We adjusted the spectra,
by subtracting the linearly decreased background and
by smoothing the noise level slightly. In the experimental spectrum,
more than twenty vibration modes were registered, and the strong
correlation was observed between our data and data of [6, 9] (see
Table), where the reasonably good berberine Raman spectra in the
range of 1000--1700 cm$^{-1}$ were presented (in [9], FT-Raman).
Unfortunately, in Tables in [6, 9], only the frequencies of the most
intense bands were given; an interpretation of vibrations was
absent, or it was incomplete or inadequate. It should be noted that
the Raman spectra of berberine obtained in [7--9] by the SERS
and SSRS methods are not applicable fully for the correct comparative
analysis of the experimental and calculated data.%\biguplus\looseness=1

In Fig. 2,{\it b}, we present the calculated Raman spectrum of a
berberine cation in the range of 1000--1700 cm$^{-1}$, by using a
scaling factor.

The frequency values and the detailed interpretation are presented in
Table. The strong correlation between experimental and calculated
spectra is observed in the range of 1000--1700 cm$^{-1}$ -- both
between frequencies and intensities of vibrations. In the range up
to 1000 cm$^{-1}$, besides a 727 cm$^{-1}$ mode, the intensities of
calculated vibrations are very small (less by 2--4 orders) in
agreement with experimental data. Notice that some modes very weak by
calculations (e.g., 1302 cm$^{-1}$) are revealed in experimental
spectra.

In Fig. 3,{\it a}, the IR-absorption spectrum of berberine
(microcrystalline powder) in the range of 800--1700 cm$^{-1}$ is
presented (after subtracting the background), about 40 rather intense
separated vibrational modes are observed, the good agreement
between our data and data of [7--9] takes place (see Table).

It should be mentioned that the obtained experimental IR-absorption and
Raman spectra appear quite similar to the IR-absorption spectra and
the FT-Raman spectrum presented in [7, 9]; but IR-frequency values were
presented in [7] only for the most intense lines, whereas they
were not presented at all in [9]. Moreover, the analysis in [7, 9] was
performed with SERS spectra that did not correspond well
to our spectra by shape. The calculated Raman and IR-absorption
spectra appear very similar to those we calculated, but differences
in frequency values are quite substantial. Experimental and
calculated Raman spectra are similar, but IR spectra are
not.\looseness=1

We attribute this discrepancy to the fact that the calculations usually
use an ion or molecule of berberine, but experiments are conducted
with commercially produced crystallohydrate. Since the dipolarity of
vibrations for IR-absorption processes is important, we have not
discounted the possibility that the presence of dipole water
molecules (and OH-groups) in specimens might have led to the
dipole-dipole interaction with berberine molecules, and the results
of the interaction might be revealed in IR-absorption spectra. The
influence on Raman spectra is weaker, because the Raman scattering
processes are related to the electron system, and the influence of
the dipole-dipole interaction on the vibrational system is less
direct.%\looseness=1

In Fig. 3,{\it b}, we present the calculated IR-absorption spectrum of
berberine (cation) in the range of 800--1700 cm$^{-1}$, by considering
a scaling factor.

The relatively strong correlation between the frequencies of
experimental and calculated IR-absorption spectra is observed in the
range of 800--1700 cm$^{-1}$; the correlation between the intensities of
vibrations is somewhat worse, for the aforementioned reasons. It
should be noted that, contrary to Raman spectra, some modes that
were calculated to be intense were not observed at all in
experimental IR-spectra, or they had only a very weak intensity
(e.g., 1149, 1242, 1285, 1408, 1471 cm$^{-1},$ {\it etc.}). In this case,
the lines of an unknown nature were
revealed in Raman and IR spectra (1127 cm$^{-1}$ (IR), 1129 cm$^{-1}$ (Raman) {\it etc.}).

\section{Conclusions}

Thus, the Raman and IR-absorption spectra have been obtained for a
berberine cation by DFT at the B3LYP/6-311++G(d,p) level. The
optimized geometry of the berberine cation was calculated as well. The
results of calculations strongly correlate with experimental data obtained for
microcrystalline berberine chloride in the region of 1000--1700 cm$^{-1}$ significant from the viewpoint of the
interaction of berberine with DNA. The obtained interpretation
of the Raman bands of berberine can be used for an analysis of
the berberine interaction with nuclei acids and other biomolecules. The
DFT method can be used, with satisfactory reliability, to calculate
vibrational spectra of other isoquinoline alkaloids of the
protoberberine group, for which obtaining the Raman
vibrational spectra would be complicated or even impossible for one
reason or another.%\looseness=1

%\vskip3mm We want to thank to S. Alekseev for essential help in
%obtaining the IR-absorption spectra, and L.~Zaika for providing the
%berberine powder.

\vskip3mm The autors are grateful to S.O. Alekseev and
M.E.~Kor\-nienko for the help in the recording of IR-spectra and to
L.A. Zaika (Institute of Molecular Biology and Genetics of the NANU)
for specimens of berberine.

%[13] M. J. Frisch, G. W. Trucks, H. B. Schlegel, G. E. Scuseria, M. A. Robb, J. R. Cheeseman, Jr., J. A. Montgomery, T. Vreven, K. N. Kudin, J. C. Burant, J. M. Millam, S. S. Iyengar, J. Tomasi, V. Barone, B. Mennucci, M. Cossi, G. Scalmani, N. Rega, G. A. Petersson, H. Nakatsuji, M. Hada, M. Ehara, K. Toyota, R. Fukuda, J. Hasegawa, M. Ishida, T. Nakajima, Y. Honda, O. Kitao, H. Nakai, M. Klene, X. Li, J. E. Knox, H. P. Hratchian, J. B. Cross, V. Bakken, C. Adamo, J. Jaramillo, R. Gomperts, R. E. Stratmann, O. Yazyev, A. J. Austin, R. Cammi, C. Pomelli, J. W. Ochterski, P. Y. Ayala, K. Morokuma, G. A. Voth, P. Salvador, J. J. Dannenberg, V. G. Zakrzewski, S. Dapprich, A. D. Daniels, M. C. Strain, O. Farkas, D. K. Malick, A. D. Rabuck, K. Raghavachari, J. B. Foresman, J. V. Ortiz, Q. Cui, A. G. Baboul, S. Clifford, J. Cioslowski, B. B. Stefanov, G. Liu, A. Liashenko, P. Piskorz, I. Komaromi, R. L. Martin, D. J. Fox, T. Keith, M. A. Al-Laham, C. Y. Peng, A. Nanayakkara, M. Challacombe, P. M. W. Gill, B. Johnson, W. Chen, M. W. Wong, C. Gonzalez, J. A. Pople,%

\rezume{%
КОЛИВАЛЬНИЙ СПЕКТР ОРГАНІЧНОЇ СПОЛУКИ\\ БЕРБЕРИНУ ТА ЙОГО
ІНТЕРПРЕТАЦІЯ\\ КВАНТОВО-МЕХАНІЧНИМ МЕТОДОМ\\ ФУНКЦІОНАЛА ГУСТИНИ }
{Н.В. Башмакова, С.Ю. Кутовий, Р.О. Жураківський,\\
  Д.М. Говорун, В.М.~Ящук}{За кімнатної температури отримано коливальні спектри (раманівський
та інфрачервоного поглинання) мікрокристалічного хлориду берберину
та проведено їхню інтерпретацію методом функціонала густини на рівні
теорії DFT B3LYP/6-311++G(d,p) в діапазоні частот 800--1700
см$^{-1}$. Спостережено добру кореляцію між експериментальними та
розрахованими частотами коливань.}

\end{document}